\documentclass[10pt,a4paper,twocolumn,aps,pra,showpacs]{revtex4-1}
\pdfoutput=1
\usepackage[utf8x]{inputenc}
\usepackage{ucs}
\usepackage{amsmath}
\usepackage{amsfonts}
\usepackage{amssymb}
\usepackage{makeidx}
\usepackage{booktabs}
\usepackage{natbib}
\usepackage{lipsum}

\usepackage{color}
\definecolor{light-gray}{gray}{0.55}

\usepackage{microtype}

\usepackage{graphicx}

\newcommand{\exv}[1]{ \langle #1 \rangle }

\newcommand{\ket}[1]{ \lvert #1 \rangle}
\newcommand{\braket}[2]{\langle #1 \vert #2 \rangle }

\newcommand{\pfrac}[2]{\frac{\partial #1}{\partial #2}}

\begin{document}

\begin{abstract}
We propose to use a time dependent imaginary potential to describe quantum mechanical tunnelling through time-varying potential barriers. We use Gamow solutions for stationary tunneling problems to justify our choice of potential, and we apply our method to describe tunneling of a mesoscopic quantum variable: the phase change across a Josephson Junction. The Josephson Junction phase variable behaves as the position coordinate of a particle moving in a tilted washboard potential, and our general solution to the motion in such a potential with a time dependent tilt reproduces a number of features associated with voltage switching in Josephson Junctions. Apart from applications as artificial atoms in quantum information studies, the Josephson Junction may serve as an electric field sensitive detector, and our studies provide a detailed understanding of how the voltage switching dynamics couples to the electric field amplitude.
\end{abstract}

\title{Effective description of tunneling in a time dependent potential with applications to voltage switching in Josephson Junctions}
\date{\today}
\author{Christian Kraglund Andersen}
\thanks{E-mail: ctc@phys.au.dk}
\author{Klaus Mølmer}
\affiliation{Department of Physics and Astronomy, Aarhus University, DK-8000 Aarhus, Denmark}


\maketitle

\section{Introduction}
Tunneling is a widely observed phenomenon in quantum mechanics and for tunneling through stationary barriers, scattering theory and good approximations based on the JWKB method are available \cite{Razavy2003}. For the general dynamical case such simple approximations do not apply and only in certain cases is it possible evaluate the tunneling rate\cite{grifoni1998driven}. The part of the wave function which is still trapped behind the barrier is evolving within the trapping region, and it may repeatedly encounter the tunnel barrier in a highly non-trivial manner. While the Schr\"odinger equation is uniquely defined in a time dependent potential, the handling of the scattering continuum components of the wave function is impeded by the demand for precise calculations on a very large interval of the coordinate variable. In this paper we propose a new, effective ansatz to handle tunneling in a time dependent potential. The method addresses tunneling in a general manner, but as a key example we will focus our attention on tunneling of a mesoscopic variable, the phase of a current-biased Josephson junction(CBJJ). 

\begin{figure}[t]
\includegraphics[width=0.95\columnwidth,clip=true,trim=3cm 8.2cm 4cm 11cm]{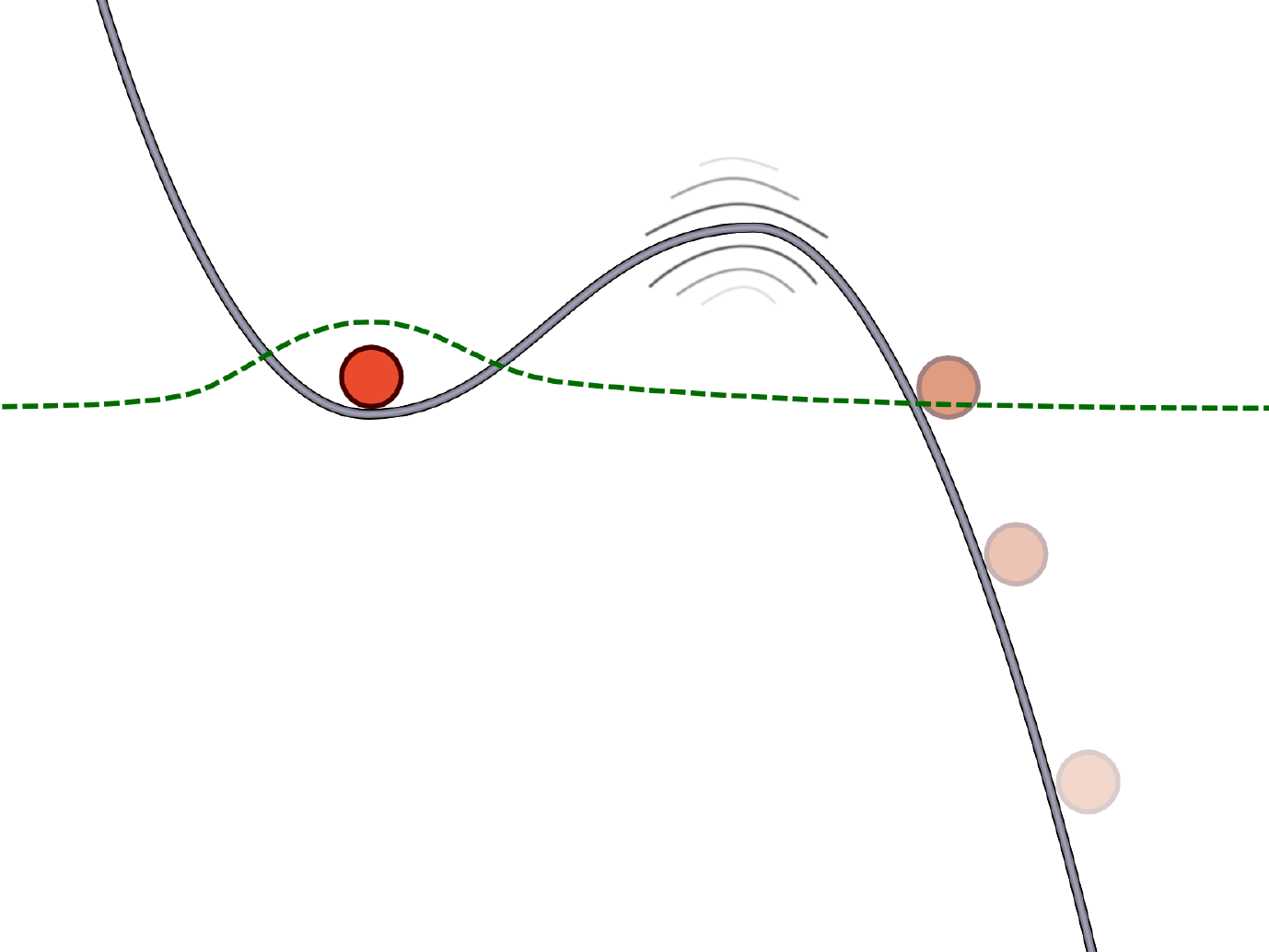}
\caption{(Color online) The superconducting phase variable in a current-biased Josephson junction behaves as a quantum particle trapped in a tilted washboard potential. The dashed (green) line is a sketch of the ground state wavefunction in one of the trapping wells. The probability that the phase particle tunnels out of the potential is increased if the barrier is modulated, e.g., by variation of the tilt or by a time dependent microwave field.} \label{fig:sketch}
\end{figure}
 
 Circuit quantum electrodynamics components have recently reached an experimental state of perfection that makes quantum interaction and control properties of these man-made devices compete with, or supersede the ones of natural, microscopic quantum systems, while at the same time retaining their fundamental interest as macroscopic quantum systems \cite{Wallraff:2004oh,PhysRevA.75.032329}. The CBJJ is a device of particular interest since the phase variable associated with the superconductor is trapped in a washboard potential, and its quantum level structure permits identification and control of pairs of states, which may be used as qubits for the realization of quantum information processing \cite{RevModPhys.73.357}. As shown in Fig.1, its motional states are subject to tunneling which corresponds to a voltage drop over the device and which may be used to distinguish the qubit states. Also, by modulation of the potential tilt or by a radiation field, a given initial state may be driven into states that undergo tunneling to running states. The switching can occur due to thermal activation \cite{PhysRevLett.93.107002} and due to macroscopic quantum tunneling. In this treatment the contribution from thermal activation will be neglected. This implies that the system can be described as an ordinary quantum particle and general methods from quantum mechanics can be applied to qualitatively and quantitatively study several features associated with its dynamics. While our approach to tunneling in a time dependent potential is general, we will here discuss its application to the CBJJ and present results analogous to observations made in CBJJ experiments. 
 
The CBJJ has also been suggested as a sensitive microwave photodetector \cite{PhysRevLett.102.173602,PhysRevA.84.063834}, and as a first step towards a low frequency quantum detection theory for such a device, we will study its response to driving by a weak microwave field.

A weak field-induced modulation, resonantly coupling the initial state to eigenstates subject to stronger tunneling, can be treated as a perturbation and for the CBJJ theory yields good agreement with experimental results \cite{PhysRevLett.55.1543, PhysRevLett.89.117901, Yu02, citeulike:10961538}. In this paper we will describe the wave-function subject to a strongly modified potential. This case is not amenable to the perturbative description, and instead, we shall formally describe the tunneling process by wave packet propagation in a time dependent imaginary potential (TDIP), such that tunneling, observed as switching of the CBJJ into the voltage state, is treated as a dynamical loss process. In this way we both determine the time dependent probability for the tunneling to occur and we appropriately describe the evolution of the wave function conditioned on no tunneling event being detected.

The paper is organized as follows.  In Sec. \ref{sec:tunnel}, we define the tunneling problem, we motivate our TDIP description, and we provide an explicit time dependent expression for the imaginary  potential applied in our numerical studies. In Sec. \ref{sec:inc_bc}, we use our method to study different regimes of tunneling dynamics with increasing bias currents applied to the CBJJ. In Sec. \ref{sec:const_bc} we fix the bias current and we analyze the performance of the Josephson junction as a field detector by examination of the switching properties under application of a driving field. Sec. \ref{sec:conclusion} concludes the paper.

\section{Tunneling of the Josephson junction phase}
\label{sec:tunnel}
\subsection{Tunneling loss and complex absorbing potentials}

The Josephson junction phase $\phi$ (see Sec.\ref{sec:dyn_cbjj}) behaves as the position of a particle described by the time dependent Schr\"odinger equation,
\begin{align}
i\hbar\pfrac{}{t} \psi(\phi,t) = H(t) \psi(\phi,t),
\end{align}
where the Hamiltonian operator $H(t)$ contains a potential and an effective kinetic energy term. The tunneling dynamics of the phase variable, which physically corresponds to the switching between different voltages across the device, is hence mathematically described as a conventional quantum tunneling problem. The purpose of this manuscript is to establish an effective, approximate theory for the tunneling dynamics in a time dependent potential, with special attention to parameter regimes relevant for the CBJJ. Following the early description by Gamow \cite{Razavy2003}, one can describe tunneling in static potentials by eigenfunctions of the time-independent Hamilton operator. If these functions are chosen with outgoing, radiating boundary conditions and used as a wave function basis, we effectively redefine the scalar product on the system Hilbert space, and the original Hamiltonian is not Hermitian. This in turn leads to the emergence of complex eigenvalues, and the loss of norm associated with the imaginary part of the energy eigenvalues represents the probability of tunnelling \cite{DelaMadrid2002,Civitarese200441,bohm1989gamow}. Effective non-Hermitian Hamiltonians also emerge in quantum optics and quantum measurement theory, where they govern the evolution of a quantum system conditioned on the absence of absorption or loss events. In these theories, the decreasing norm of the state vector provides the probability for the evolution to occur without these events happening, while one may simulate the complete dynamics including random detection events by suitable application of "quantum jump" operators \cite{RevModPhys.70.101,wiseman2010quantum,PhysRevLett.68.580}.

At a given time the tunneling probability is governed by the wave function obtained by propagation with our non-Hermitian Hamiltonian, \textit{i.e.}, the state conditioned on the detection of no previous switching events. The amount of trapped population and its actual wave function therefore yields the local loss rate due to tunneling and the whole time dependent dynamics should be well described by the Gamow vectors and their complex eigenvalues. Another approach to yield complex eigenvalues is by directly implementing a non-Hermitian Hamiltonian with respect to the original Hilbert space of square integrable wave functions, for example by the introduction of a complex absorbing potential (CAP). A properly designed CAP may thus describe the same essential physics as the Gamov vectors, i.e., the evolution of the unnormalized states may yield identical or very similar wave function behaviour in the spatial and temporal range of interest \cite{sommerfeld1999temporary}.

There is a rich literature \cite{Muga2004357} on the identification of suitable CAPs, but since we shall be dealing with the further complexity of tunneling through a time dependent potential, we shall merely propose a simple, physically motivated \emph{ansatz} for our time dependent imaginary potential (TDIP), $V_{im}(\phi,t)$, and solve the time-dependent Schrödinger equation
\begin{align}
i\hbar\pfrac{}{t} \psi(\phi,t) = H(t) \psi(\phi,t) - iV_{im}(\phi,t) \psi(\phi,t).
\end{align}

\subsection{The dynamics of a CBJJ}
\label{sec:dyn_cbjj}
In this section we outline the particular properties and the effective quantum description of a CBJJ by a one-dimensional Schr\"odinger equation for the phase variable $\phi$. The Josephson junction is composed of two superconductors, R(ight) and L(eft), separated by a thin isolating layer. The macroscopic wave functions of R and L, $\psi_{R,L}$ differ by a phase factor $\psi_L = e^{i\phi} \psi_R$, which constitutes the macroscopic quantum degree of freedom of the system.  This phase can be described \cite{devoret2004} as a quantum particle with mass $M = C(\Phi_0/2\pi)^2$, where $C$ is the capacitance of the junction and $\Phi_0 = \frac{h}{2e}$ is the flux quantum.  The particle experiences a potential
\begin{align}
U_0(\phi) = -E_J( I \phi + \cos \phi )
\end{align}
where $E_J = \frac{I_c \Phi_0}{2\pi}$ is the Josephson energy and $I = \frac{I_b}{I_c}$, with $I_b$ being the bias current applied to the junction. When $I_b$ exceeds the critical current $I_c$, the potential tilt dominates the harmonic variation with $\phi$, and the phase becomes classically unbound.

We further include the interaction of the CBJJ with a time dependent microwave field via the potential term,
\begin{align}
U_{mw} = -E_J \eta \phi \sin{\Big(\omega_{mw} t\Big)}
\end{align}
where $\eta$ is proportional to the field amplitude.

The CBJJ is thus described by the time-dependent Schrödinger equation
\begin{align}
 i \hbar \pfrac{ }{t}\psi(\phi,t) =& H(t) \, \psi(\phi,t) \nonumber \\=& \Big(- \frac{\hbar^2}{2M} \pfrac{^2}{\phi^2}  + U(\phi,t)  \Big) \psi(\phi,t) \label{eq:Schr}
\end{align}
with $U(\phi,t) = U_0 + U_{mw}$. This time dependent potential is sketched in Fig. 1, indicating also the tunneling process, responsible for the switching of the Josephson junction.

\subsection{Outgoing states and absorbing potential}
An imaginary absorbing potential, extending beyond the outer turning point of a potential barrier would seem a natural candidate to remove the tail of the wave function as tunneling develops. Since we want the potential to remove the projection of our wave function on the running states in that region, it is useful to apply approximate solutions for their position (phase) and time dependence in our ansatz for the TDIP. From \cite{PhysRevB.72.134528} we have an approximate expression for the running state of the Josephson junction phase variable $\phi > \phi_{turn}$, initially at time $\tau$ in the ground state, in the linear potential region beyond the outer classical turning point of the potential barrier $\phi_{turn}$,
\begin{align}
\psi_{out}(\phi,t,\tau) = &\sqrt{\frac{\Gamma}{\sqrt{2\phi'}\Phi_0 \omega_p(\tau)}} e^{i \frac{\hbar \omega_p {\phi'}^{3/2}}{6\sqrt{2} E_J}} \nonumber\\
&\times e^{- \big( i \frac{\omega}{\omega_p} + \frac{\Gamma}{2\omega_p} \big) ((t-\tau)\omega_p - \sqrt{2\phi'})} \label{eq:runningstate}
\end{align}
with $\phi' = \phi - \phi_0$, where $\phi_0$ is the initial equilibrium position (bottom of the well) and $\hbar\omega$ is the energy difference between the bottom of the well and the energy of the lowest quasi-bound state. We have further introduced the frequency parameter $\omega_0 = \sqrt{\frac{2\pi I_c}{C\Phi_0}}$ and the plasma frequency $\omega_p = \omega_0 (1 - I^2)^{1/4}$. The wavefunction (\ref{eq:runningstate}) is defined for  $t\omega_p > \sqrt{2\phi}$, and we assume $I < 1$. The rate parameter $\Gamma$ in \eqref{eq:runningstate} which attains a definite value in a static potential, will be briefly discussed below.

Our aim is to remove the projection of our time dependent wave function $|\psi(t)\rangle$ on a suitable set of running states of the form  \eqref{eq:runningstate}, $P_{out} \ket{\psi} = \sum_{\psi_{out}} \ket{\psi_{out}} \braket{\psi_{out}}{\psi}$. We model this operation by a time-dependent complex potential, obtained as an integral over different running states, emitted within the past interval of time $\Delta t$,
\begin{align}
V_{im} (\phi,t) = \, \beta \int_{t-\Delta t}^t |\psi_{out}(\phi,t,t')|^2 \, dt'.\label{eq:time-int}
\end{align}
The parameter $\beta$ serves both as an adjustable strength parameter and as a convenient normalization for the temporal integration.
We assume a constant value of $\Gamma$ and $\Delta t$, with $\Delta t^{-1} \ll \Gamma \ll \omega_p, $, and we evaluate the integral \eqref{eq:time-int} only for values $\phi$ larger than the outer turning point, $\phi_{turn}$, of the potential.  The dependence on $\Gamma$ drops out, and we obtain
\begin{align}
V_{im} (\phi,t) = \begin{cases}  \frac{\beta}{\sqrt{2\phi} \, \Phi_0 \omega_p(t)} & \text{ for } \phi > \phi_{turn}(t) \\ 0 & \text{ for } \phi < \phi_{turn}(t).  \end{cases} \label{eq:vim}
\end{align}
Here, we recall the time-dependence of the plasma frequency as the effective bias-current is changing with time.

Unlike normal CAPs used in time-independent problems, this potential is modelled to absorb the running state components \eqref{eq:runningstate} pertaining to the time dependent Hamiltonian, and our ansatz imaginary potential indeed attains finite values only beyond the time dependent outer classical turning point of the real potential. We emphasize that (\ref{eq:vim}) is only an ansatz, but it ensures that the depletion of the wave function norm is properly associated with the probability that the particle has tunnelled through the barrier.

While it is generally a challenge to design CAPs that do not reflect part of the wave packets impinging on them, our application only concerns the small wave function amplitude that has tunneled through the real potential barrier, and we have verified that reflection is insignificant in our calculations. To account for the running solution, the TDIP must furthermore suppress the wave function before it hits and gets reflected by the border of the grid used for the calculation, and a sufficiently large grid is readily identified in the numerical calculations.

\subsection{Friction and junction resistance}
We have now presented our candidate TDIP, but before we proceed to numerical examples let us include the effects of friction (junction resistance). Methods have been developed to incorporate friction and diffusion in the quantum mechanical description of Brownian motion of a particle coupled to its environment, \cite{strunz2001brownian,PhysRevA.66.052105}.  Assuming zero temperature and Markovian noise correlation in the environment, we can describe the effects of the junction resistance, $R$, by a non-linear imaginary potential term, $- i\zeta \Big(\phi - \exv{\phi}_t\Big)^2$ , where $\zeta \propto \frac{1}{RC}$ and $ \exv{\phi}_t = \int_{-\infty}^{\infty} \phi \, |\psi(\phi,t)|^2 \, d\phi$ is the mean value of $\phi$ at time $t$. When added to the Hamiltonian this term penalizes large variations of $\phi$ around its mean and decoheres the spatial wave function, which in turn leads to friction for the phase variable $\phi$. It also leads to a loss of norm, albeit on a typically slower scale than the tunneling dynamics. We shall include this term in our calculations on the CBJJ, but we shall renormalize the wave function with respect to the loss it incurs, so that we unambiguously associate the wave function loss of norm with the tunneling dynamics.

To summarize, we treat the entire problem by solving the non-linear Schrödinger equation
\begin{align}
 i \hbar \pfrac{ }{t}\psi(\phi,t) =& \,\Big(- \frac{\hbar^2}{2M} \pfrac{^2}{\phi^2}  + U(\phi,t)  \Big) \psi(\phi,t)  \nonumber \\
 &\phantom{\,\Big(}  - iV_{im}(\phi,t) \, \psi(\phi,t) \nonumber \\
 &\phantom{\,\Big(} - i\zeta \Big(\phi - \exv{\phi}_t\Big)^2 \psi(\phi,t) \label{eq:schreq}
\end{align}
and renormalize with respect to the loss of norm caused by the last term.

\begin{figure}[t]
\includegraphics[width=0.95\columnwidth,clip=true,trim=2cm 15cm 2cm 1.6cm]{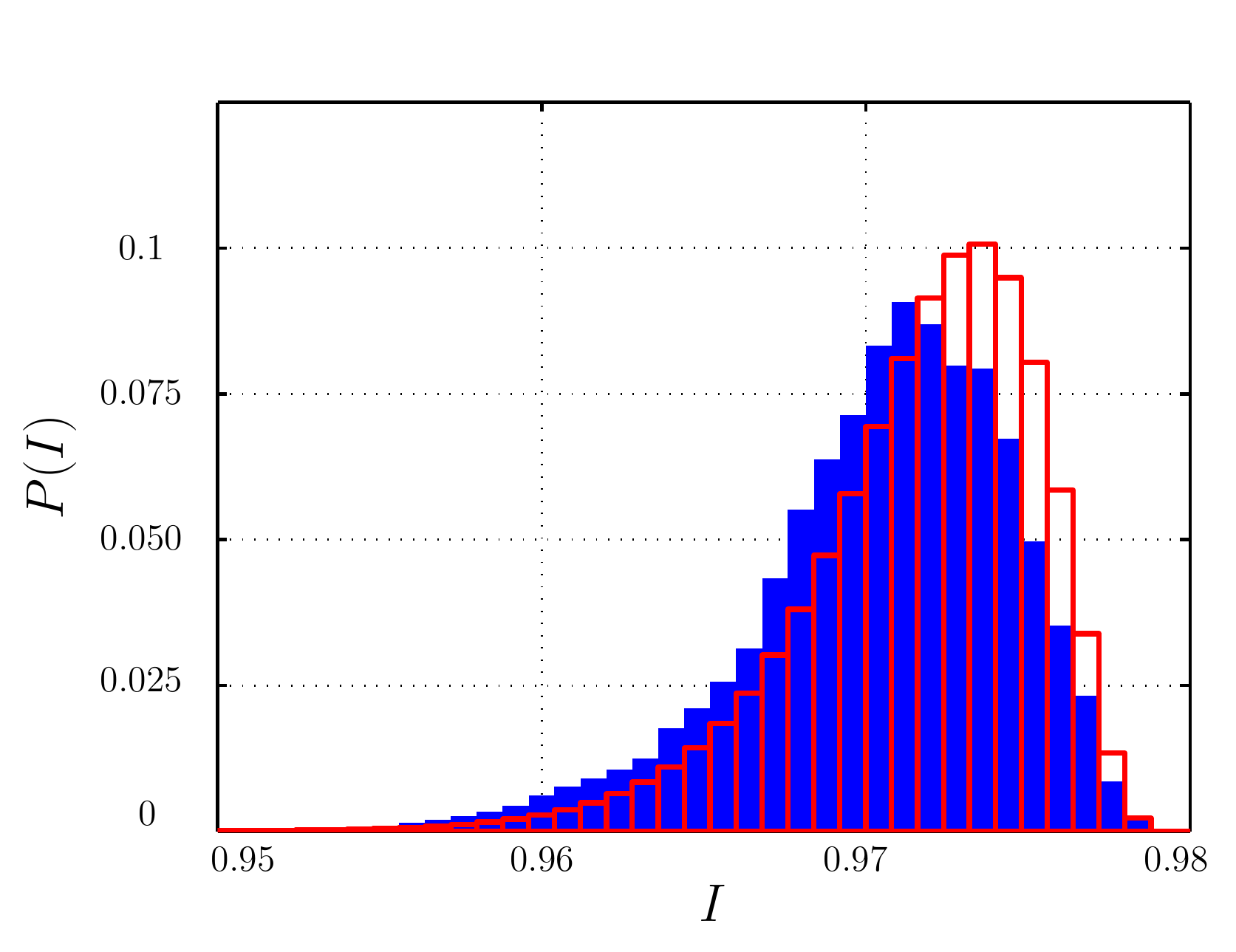}
\caption{(Color online) Blue (solid) bars show the switching current distribution simulated with an absorbing potential. Red (hollow) bars are calculated by Eq. \eqref{eq:CL}. We have chosen $C$ and $R$ such that $\omega_0 = 0.0183 \; E_J/\hbar$ and $\zeta = 8.4 \times 10^{-4} E_J$. The bias current is increased linearly from $I=0.2$ at a slow rate. $dI/dt = \frac{5 E_J}{6 \hbar} $. Results are shown for $I>0.95$.} \label{fig:compare}
\end{figure}

\begin{figure*}[t]
\begin{minipage}[t]{0.49\linewidth}
\includegraphics[width=0.95\textwidth,clip=true,trim=2cm 20cm 2cm 1.6cm]{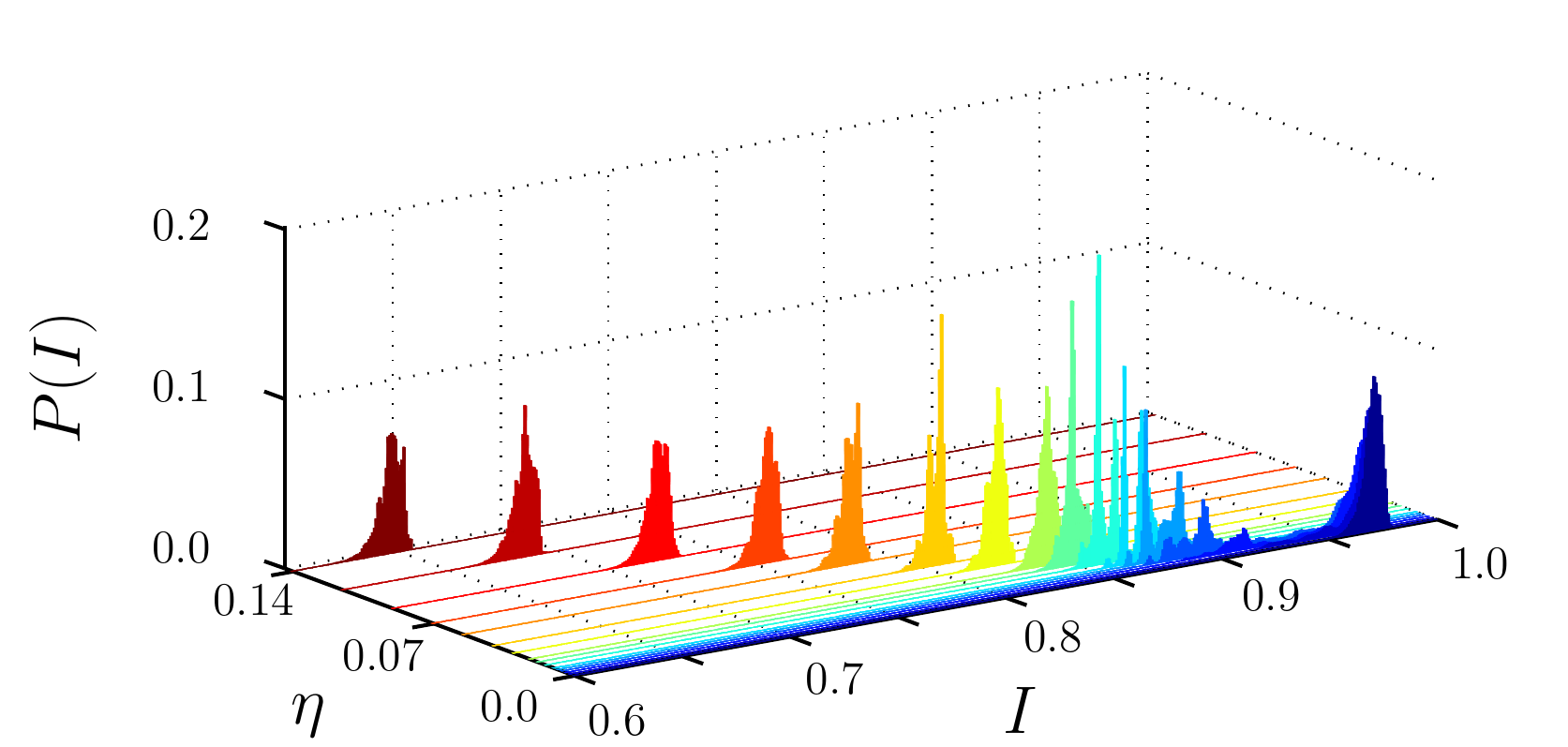}
(a1)

\includegraphics[width=0.95\textwidth,clip=true,trim=2cm 19cm 2cm 1.6cm]{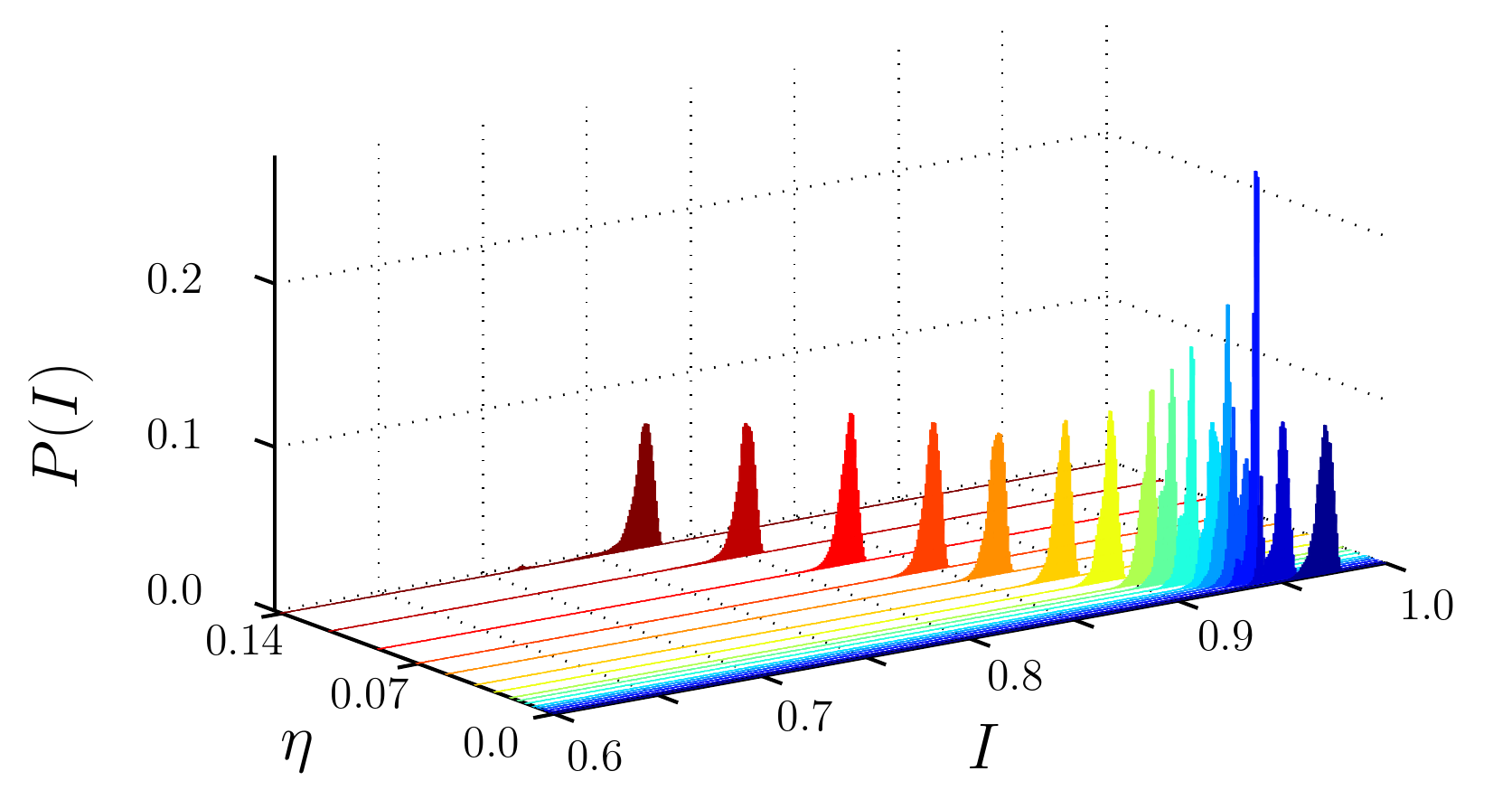}
(b1)
\end{minipage}
\begin{minipage}[t]{0.49\linewidth}
\includegraphics[width=0.95\textwidth,clip=true,trim=2cm 20cm 2cm 1.6cm]{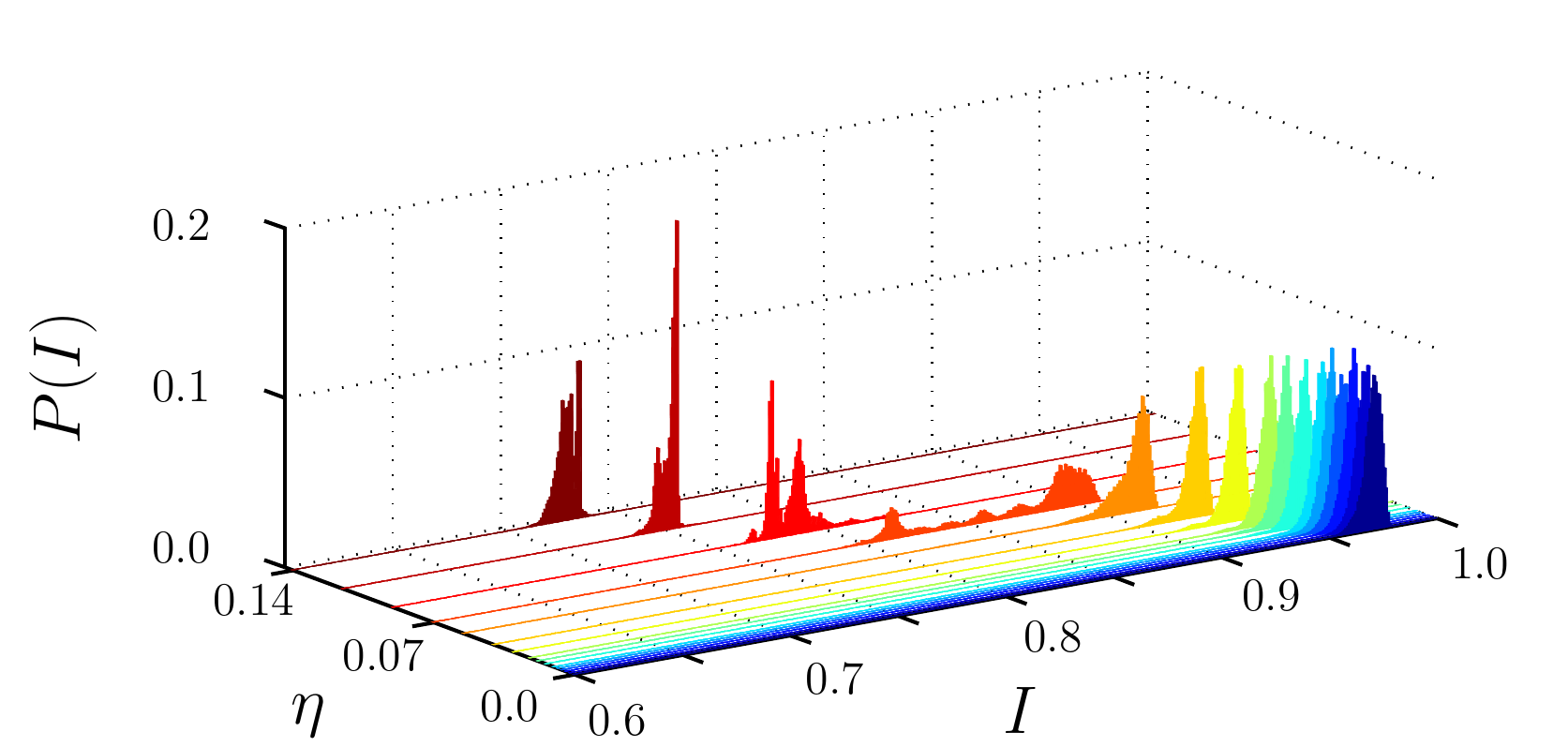}
(c1)

\includegraphics[width=0.95\textwidth,clip=true,trim=2cm 19cm 2cm 1.6cm]{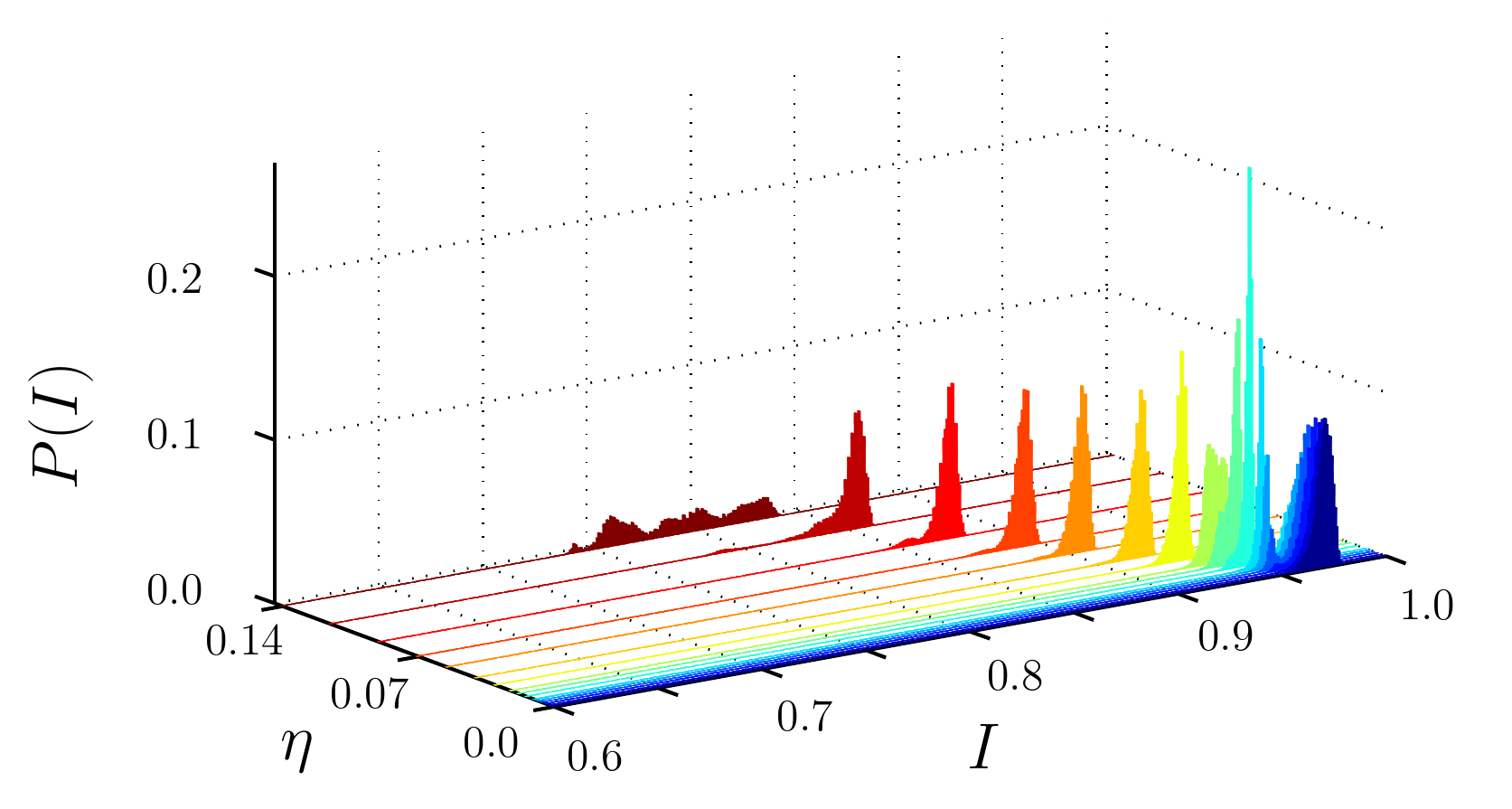}
(d1)
\end{minipage}

\vspace{0.5cm}

\includegraphics[width=0.3205\textwidth,clip=true,trim=1.35cm 19cm 8.8cm 1.5cm]{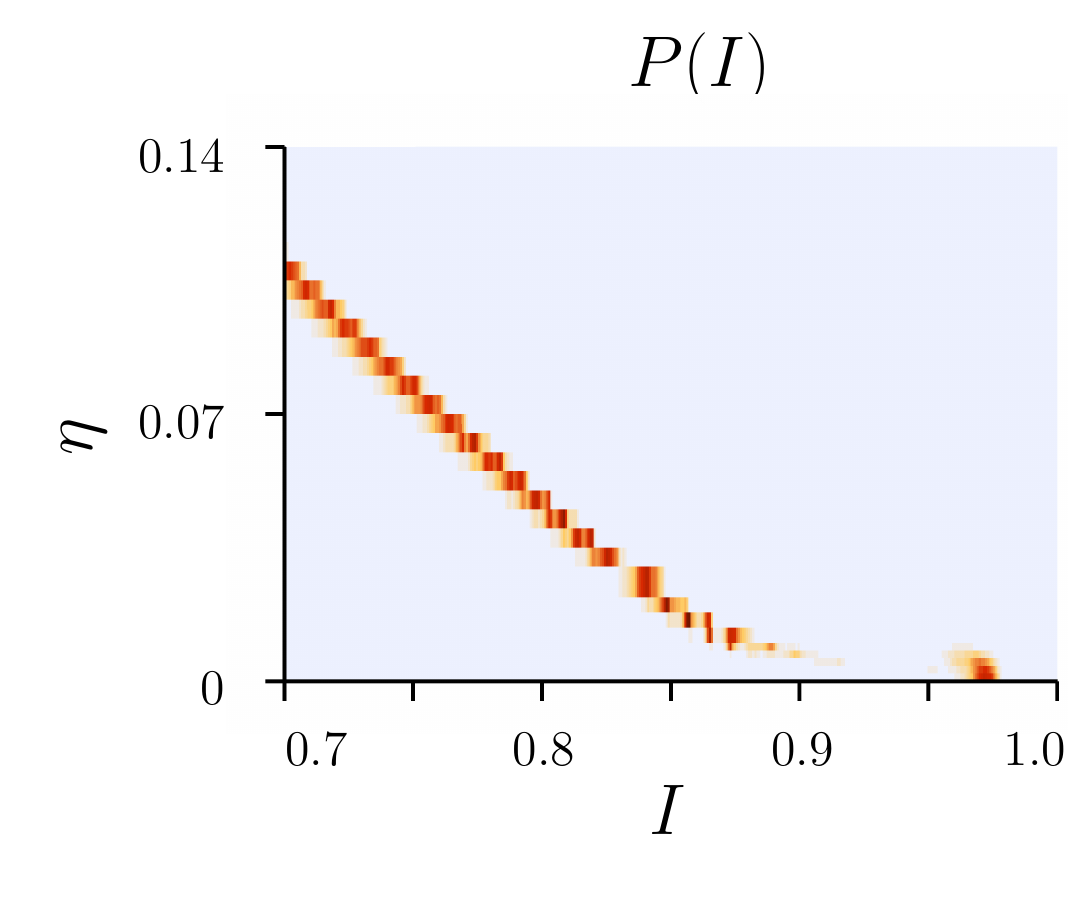}\,
\includegraphics[width=0.1675\textwidth,clip=true,trim=4.12cm 19cm 11.2cm 1.5cm]{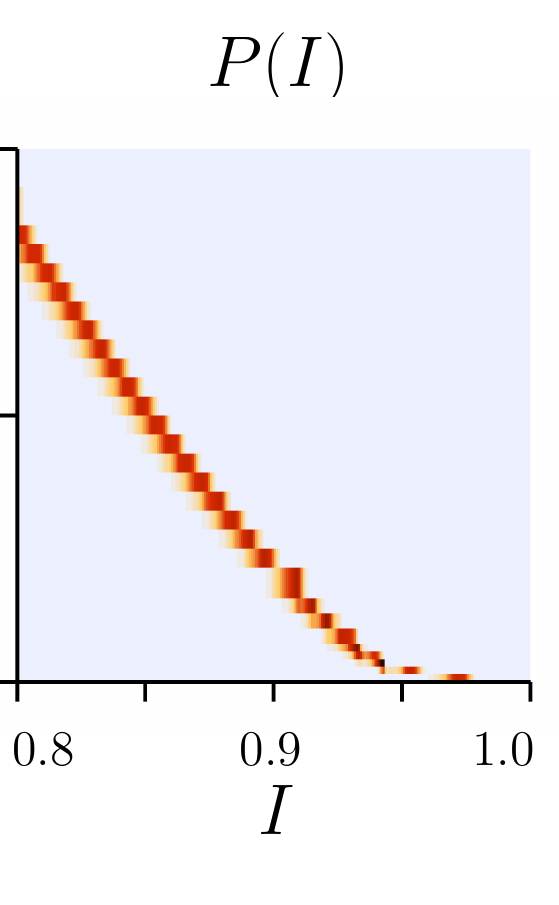}
\includegraphics[width=0.238\textwidth,clip=true,trim=4.08cm 19cm 8.8cm 1.5cm]{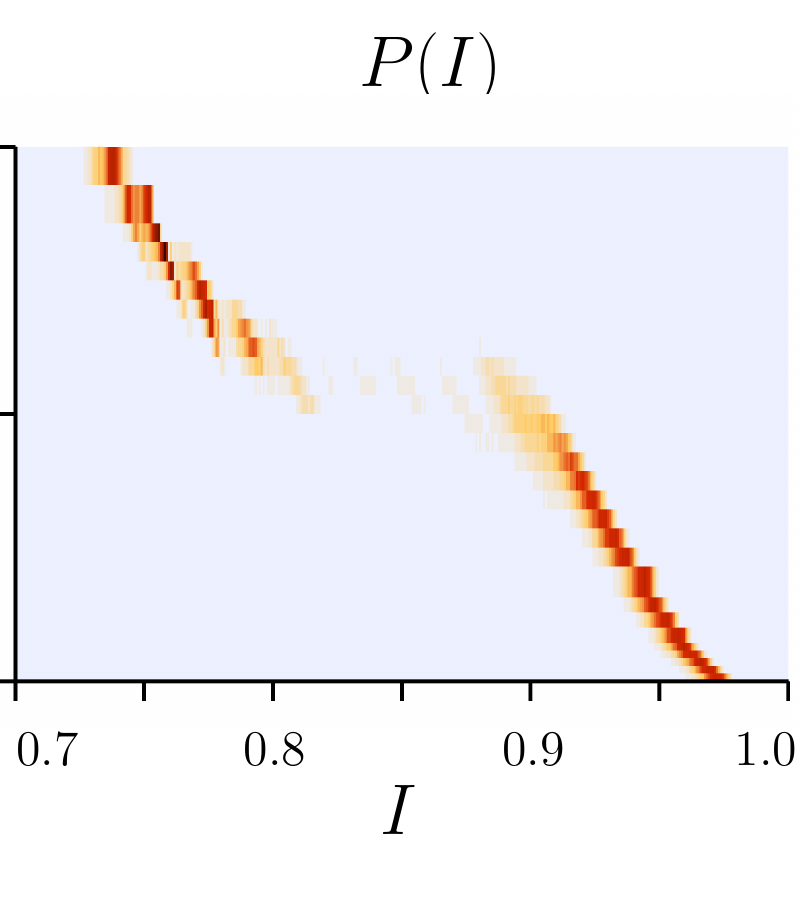}\,
\includegraphics[width=0.246\textwidth,clip=true,trim=4.12cm 19cm 8.5cm 1.5cm]{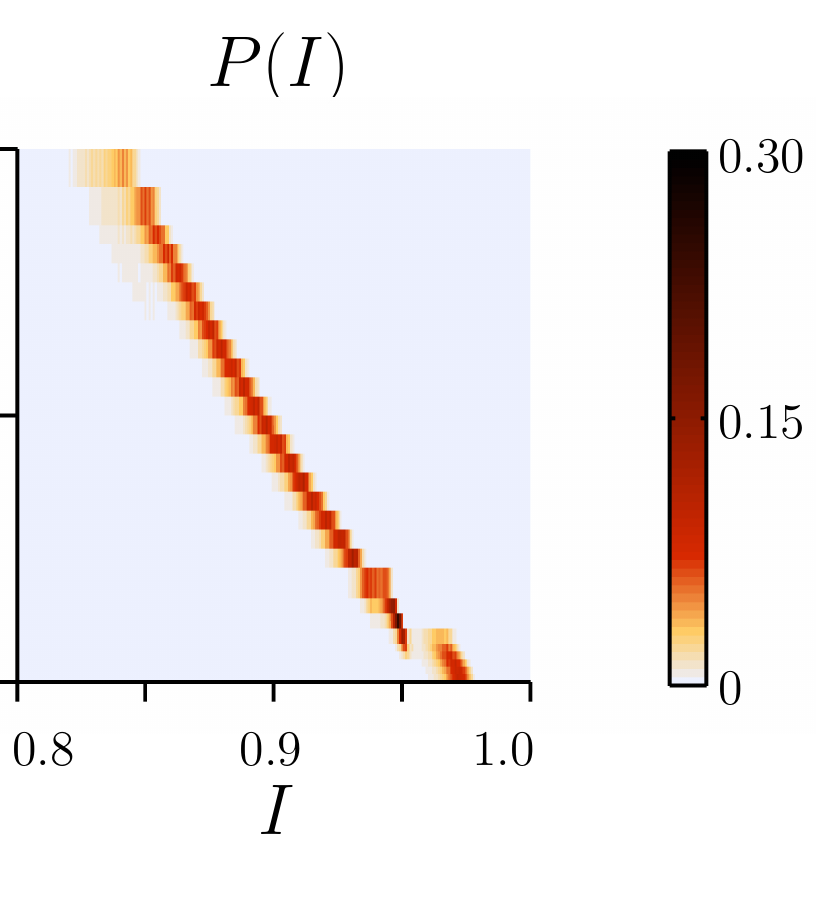}

\begin{minipage}{0.094\linewidth}
\,
\end{minipage}
\begin{minipage}{0.238\linewidth}
(a2)
\end{minipage}
\begin{minipage}{0.1675\linewidth}
(b2)
\end{minipage}
\begin{minipage}{0.238\linewidth}
(c2)
\end{minipage}
\begin{minipage}{0.1675\linewidth}
(d2)
\end{minipage}\begin{minipage}{0.094\linewidth}
\,
\end{minipage}

\caption{(Color online) Switching current distributions for a CBJJ driven by a field. In all figures we have chosen $C$ and $R$ such that $\omega_0 = 0.0183 \; E_J/\hbar$ and $\zeta = 8.4 \times 10^{-4} E_J$. The frequencies and hence the resonance currents are chosen as $\omega_{mw} = 0.64 \omega_0$, $I=0.90$ in (a1) and (a2), $\omega_{mw} = 0.51 \omega_0$, $I=0.95$ in (b1) and (b2), $\omega_{mw} = 0.36 \omega_0$, $I=0.98$ in (c1) and (c2), and $\omega_{mw} = 0.255 \omega_0$. $I=0.987$ in (d1) and (d2). In panels (c1-2) and (d1-2) the resonance current is above the current of the primary zero-field peak. In (a1-d1) the applied microwave amplitude attains the values, $\eta = $ 0, 0.002, 0.004, 0.006, 0.008, 0.010, 0.014, 0.018, 0.022, 0.030, 0.040, 0.055, 0.070, 0.090, 0.115, 0.140, which yield the curves shown in the front (dark-blue) towards the back (dark-red) in each panel. In (a2-d2) we see the color plots showing the same result as the three dimensional plots in (a1-d1). The calculations are carried out with a bias current that increases linearly with time from $I=0.2$ at a rate equal to $dI/dt = \frac{5 E_J}{6 \hbar} $. (In (a1-d1) the results are only shown for $I > 0.6$, since $P(I) \ll 1$ for the smaller bias currents. In (a2) and (c2) we show the results for $I > 0.7$, while in (b2) and (d2) the results are shown for $I > 0.8$.)} \label{fig:switch_omega}

\end{figure*}

\begin{figure}[t]
\includegraphics[width=0.95\columnwidth,clip=true,trim=2cm 15cm 2cm 1.6cm]{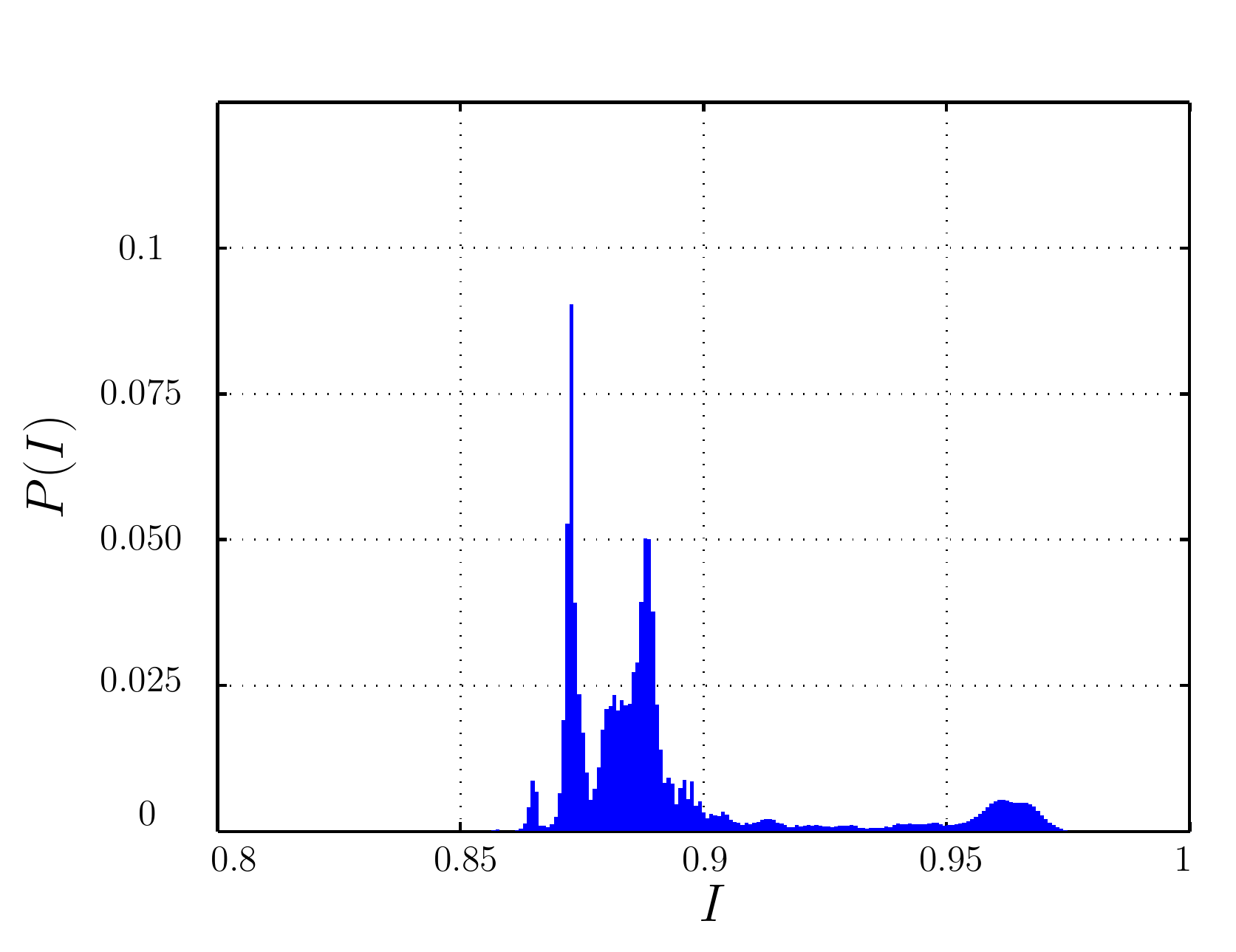}
\caption{(Color online) Switching current distributions for a CBJJ driven by a field. The parameters are chosen as in Fig. \ref{fig:switch_omega} (a) and with $\eta = 0.008$. These parameters give rise to a multi-peak structure of the switching current distribution.} \label{fig:multi}
\end{figure}

\subsection{Tunneling rates and and switching current distributions}

In a time-independent potential, the tunneling rate of the Josephson junction phase variable has been determined by
Caldeira and Leggett \cite{PhysRevLett.46.211,Caldeira1983374},
\begin{align}
\gamma_{CL} = \frac{\omega_p}{2\pi} \sqrt{\frac{120\pi\cdot7.2  \Delta U}{\hbar \omega_p}} e^{-\frac{7.2\Delta U}{\hbar \omega_p}\big(1+ \frac{0.87}{\omega_p R C}\big)} \label{eq:CL}
\end{align}
where $\Delta U = 2E_J \big(\sqrt{1-I^2} - I \text{acos}(I)\big)$ is the barrier-height, and where effects due to the friction are also taken explicitly into account.  This analytic expression available for CBJJs qualify them as ideal candidates to test our general tunneling description.

The gradually decreasing norm, $||\psi(t)||^2$ of our numerically determined wave packet is interpreted as the probability that the phase variable has not tunneled until time $t$. The probability for a current switching event in the next infinitesimal time interval $dt$, is thus simply given by the loss of norm in that interval, and conditioned on no previous event, the switching rate of the CBJJ reads
\begin{align}
\gamma_t = -\frac{d||\psi(t)||^2}{dt}/||\psi(t)||^2.
\end{align}

It is convenient in experiments to determine the switching current distribution, i.e., the probability distribution $P(I)$ for switching events to occur at different values of the bias current $I$, while this is being ramped up slowly, $I(t)=I_0 + \frac{dI}{dt}\cdot t$.

Our wave packet propagation yields the surviving (non-switching) population $||\psi(t)||^2$ as function of time, and we directly obtain the corresponding switching current distribution, obeying
\begin{align}
P(I)\Delta I &= -\frac{d||\psi(t)||^2}{dt} \Delta t  \nonumber \\ &= -\frac{d||\psi(t)||^2}{dt} \Big(\frac{dI}{dt}\Big)^{-1} \Delta I, \label{eq:dist}
\end{align}
evaluated at the time $t$ such that $I=I(t)$ and $\Delta t$ being the infinitesimal time interval in which $I$ increases by 
$\Delta I$.

Under the assumption of a slowly ramped bias current, the rate $\gamma_{CL}$ found by Caldeira and Leggett leads to a switching current density, expressed as a product of the current tunneling rate with the survival probability until the value $I$ is reached during the ramp,
\begin{align}
P_{CL}(I) = \Big(\frac{dI}{dt} \Big)^{-1}\gamma_{CL}(I) \, e^{-\int_{I_0}^{I} \big(\frac{dI}{dt} \big\vert_{I=I'} \big)^{-1} \gamma_{CL}(I')dI'} . \label{eq:cldist}
\end{align}

In Fig. \ref{fig:compare} we see, that our calculation, matches the result of the quasi-static switching current distribution, \eqref{eq:cldist}, reasonably well as a function of the bias current.

\begin{figure*}[t]
\includegraphics[width=0.95\columnwidth,clip=true,trim=2cm 18.2cm 2cm 1.4cm]{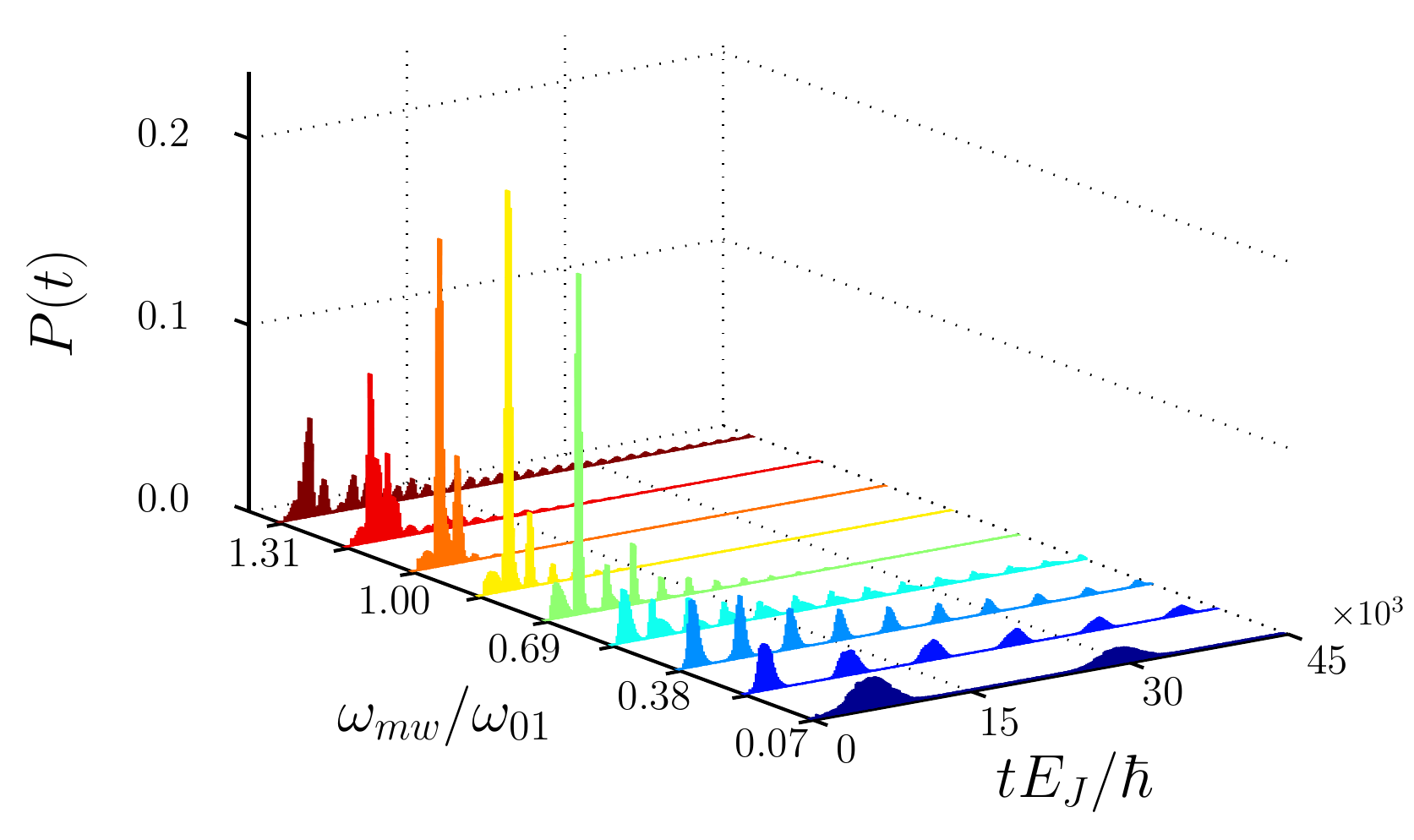}\hspace{1cm}
\includegraphics[width=0.95\columnwidth,clip=true,trim=2cm 18.2cm 2cm 1.4cm]{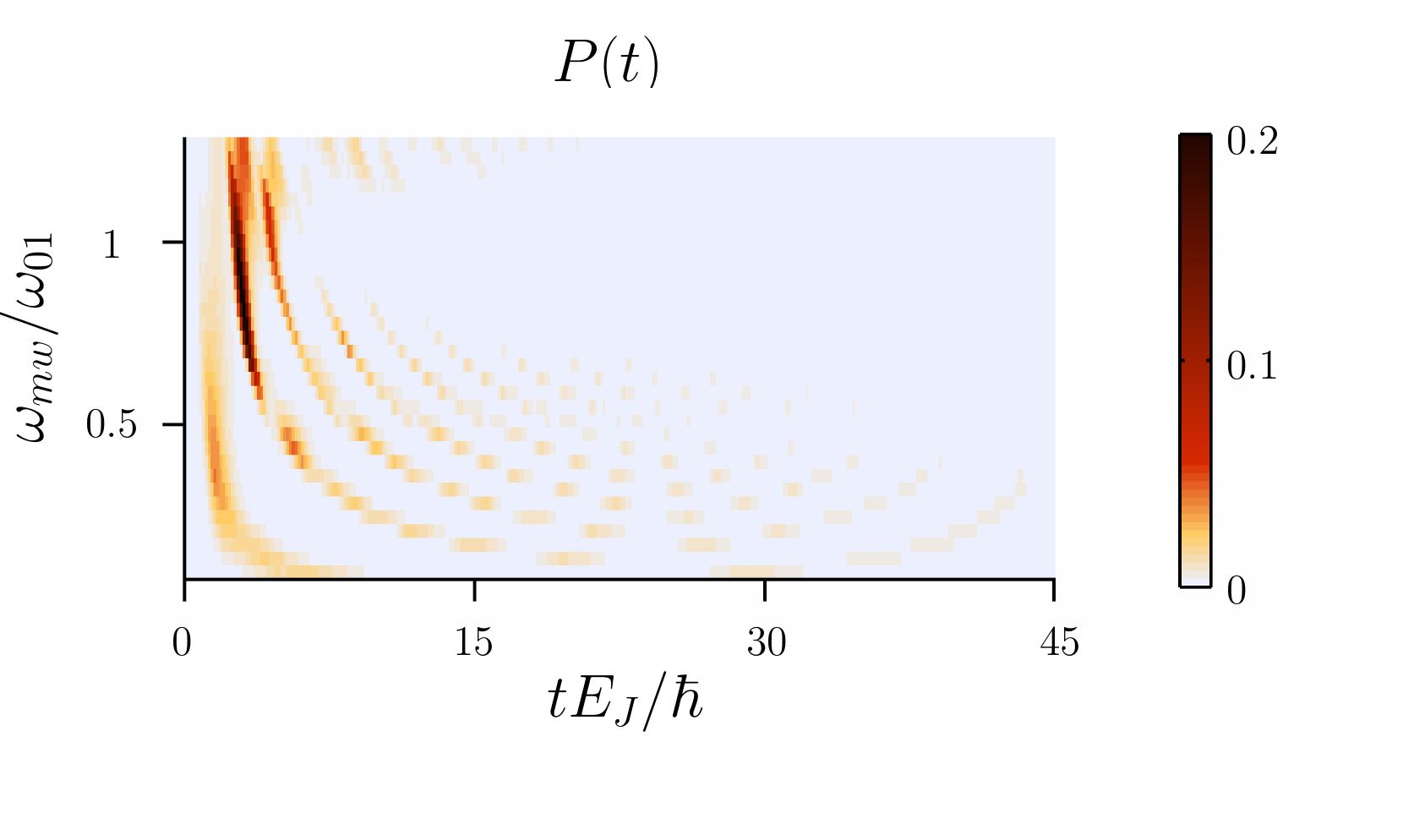}
\caption{(Color online) Switching distributions for different frequencies with a fixed bias current at $I=0.96$. We have chosen $C$ and $R$ such that $\omega_0 = 0.0183 \; E_J/\hbar$ and $\zeta = 8.4 \times 10^{-4} E_J$. This gives a zero-field energy splitting between the ground and first exited states at $\omega_{01} = 0.48 \omega_0$. To the left, we have increasing frequency from the front (dark-blue) to the back (dark-red) at $\omega_{mw} = ($0.07, 0.23, 0.38, 0.54, 0.69, 0.84, 1.00, 1.15, 1.31$) \omega_{01}$. To the right, we show a color plot of the same switching distributions where the signal from later cycles decrease and eventually disapper. The power of the field is chosen with $\eta = 0.020$.} \label{fig:switch_frequency}
\end{figure*}

\begin{figure*}[t]
\includegraphics[width=0.95\columnwidth,clip=true,trim=2cm 19cm 2cm 1.6cm]{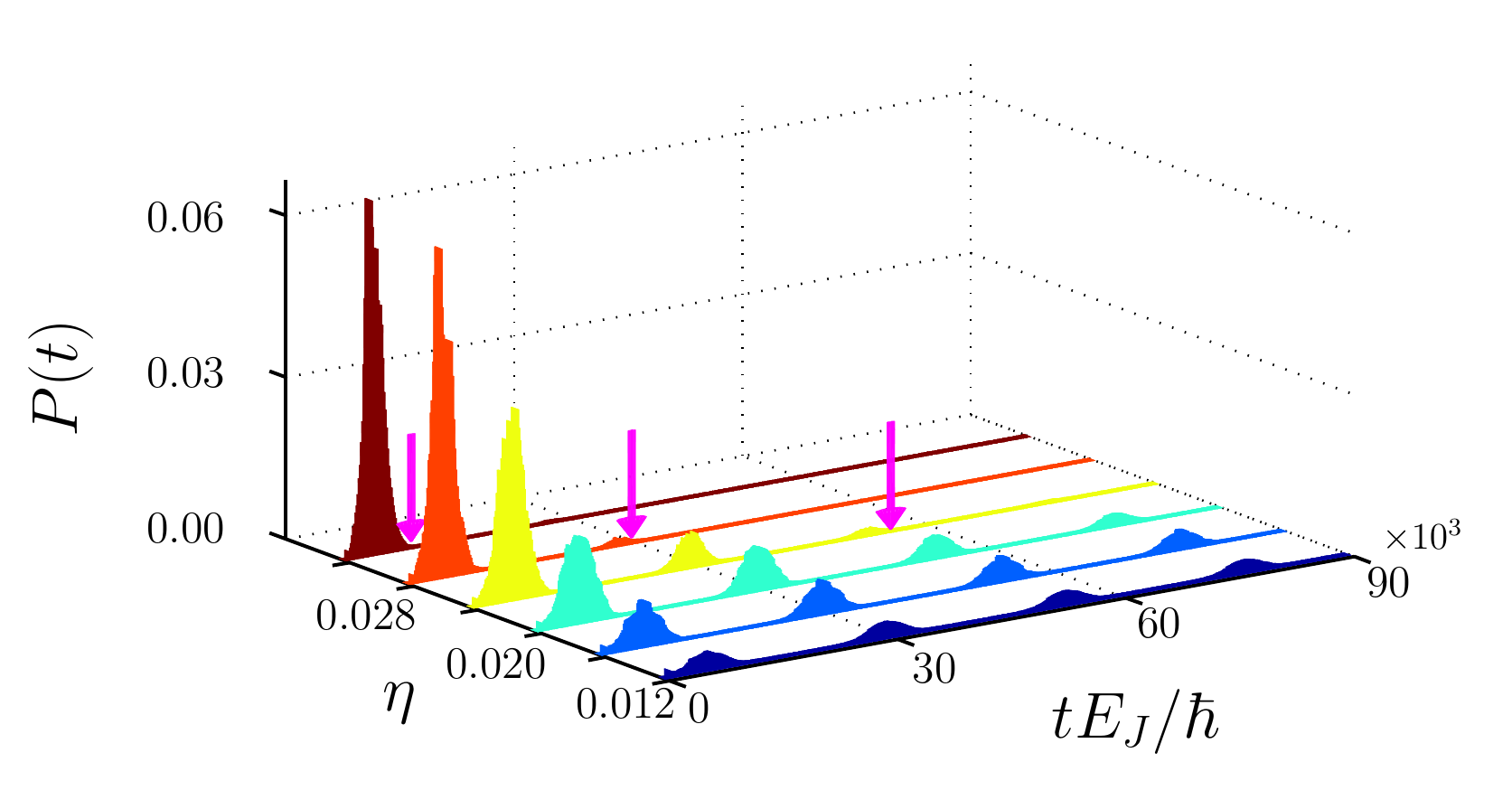}\hspace{1cm}
\includegraphics[width=0.95\columnwidth,clip=true,trim=2cm 18.2cm 2cm 1.4cm]{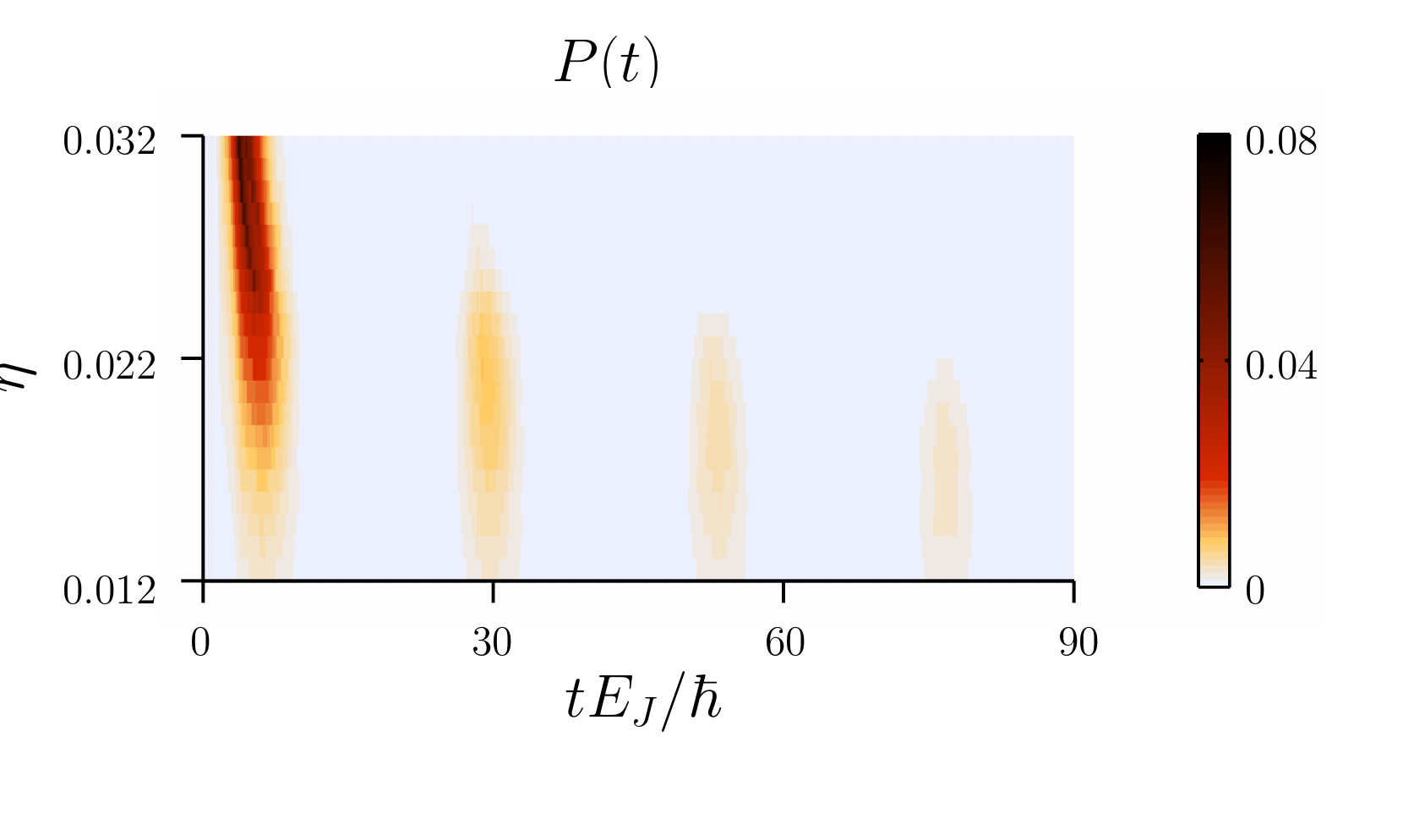}
\caption{(Color online) Switching distributions for different amplitudes with a fixed bias current at $I=0.96$. We have chosen $C$ and $R$ such that $\omega_0 = 0.0183 \; E_J/\hbar$, $\omega_{01} = 0.48 \omega_0$ and $\zeta = 8.4 \times 10^{-4} E_J$. We drive the junction at a very low frequency $\omega_{mw} = 0.036 \omega_0$. To the left the power of the field is increasing from the front (dark-blue) to the back (dark-red) at $\eta = $ 0.012, 0.016, 0.020, 0.024, 0.028 and 0.032. The arrows (magenta) indicate when the accumulated probability for a switch equals unity. In the strong field case this happens after only one half cycle. To the right we show a color plot of the same switching distributions as a function of the driving field the amplitude.} \label{fig:switch_amplitude}
\end{figure*}

\section{Driving with Increasing Bias Current}
\label{sec:inc_bc}
To study a system with a more complicated time-dependent tunneling dynamics we now include a microwave field with a constant power (constant $\eta$) and a frequency $\omega_{mw}$, while we increase $I$ with a low constant rate. Many experimental results are available for the CBJJ under such conditions, and this special example of tunneling dynamics serves as the primary application of our theoretical description.

The evolution of the system is calculated numerically using equation \eqref{eq:schreq} with the imaginary potential given in \eqref{eq:vim}. The parameters are chosen to represent a Josephson junction with a critical current of ${\sim}2$ A. The results of the simulations are shown in Fig. \ref{fig:switch_omega}(a1-d1) where the probability distribution is plotted as three dimensional plot, while we in \ref{fig:switch_omega}(a2-d2) plot the switching current probability distributions as two dimensional color plots to give both a qualitative and a quantitative overview of the results. In the following subsections we will separately discuss many of the features observed in Fig.3.

\subsection{Resonance Peaks}

We notice in Fig. \ref{fig:switch_omega}(a1-2) and \ref{fig:switch_omega}(b1-2) that when the microwave field strength is increased, in addition to the conventional switching current peak a second peak appears. In the two dimensional color plots this is seen as the opening of a gap in the switching current curve. This second peak is associated with the tunnelling via the excited phase eigenstate \cite{PhysRevB.81.144518,PhysRevB.77.104531}, and hence the peak starts appearing where the frequency $\omega_{mw} = \omega_{01}(I)$, with the resonance frequency 
\begin{align}
\omega_{01}(I) = \omega_p(I) \Big(1 - \frac{5\hbar \omega_p(I)}{36 \Delta U(I)} \Big).
\end{align}

In Fig. \ref{fig:switch_omega}(d1-2) at a much higher field strength we observe a second peak at the same bias current as in Fig. \ref{fig:switch_omega}(b1-2). But, since, in this panel, the microwave frequency is half the resonance frequency, the excitation of the phase variable is a second order process \cite{PhysRevLett.90.037003}.

When the power is further increased the peaks move linearly to lower bias current, which is also well-understood \cite{PhysRevB.68.060504}. This behaviour is clearly observed in Fig. \ref{fig:switch_omega}(a2-d2). The strong microwave field then effectively suppresses the potential barrier as the power is increased.

\subsection{Multi-peak Structure}
In Fig. \ref{fig:switch_omega}(a), after the revival of the second peak, a multi-peak structure appears. This is shown in Fig. \ref{fig:multi}. Since we only drive the junction with a single frequency field, these extra peaks cannot be understood as conventional resonances. They are, however, a well known effect associated with the multi-peaked Fourier transform of a frequency chirped electric field amplitude \cite{Davis06freq}, see also \cite{PhysRevLett.81.2679}. The analysis in \cite{Davis06freq} leads to peaks with higher density for the largest detunings and, although the correspondence between the two physical problems is only approximate, it is consistent with our observation that the multi-peak structure is not discernible in Fig. \ref{fig:switch_omega}(b). We should emphasize that in typical experiments, $\frac{dI}{dt}$ is smaller than in our calculations leading to more narrow peaks than seen here.

\subsection{Dynamical Bifurcation}
In Fig. \ref{fig:switch_omega}(c1-2) we see yet another interesting feature of a driven CBJJ. The resonance condition lies here at a bias current above the zero-field peak current, which implies that we see no resonance peak. However for very strong fields we observe a very broad splitting into a multi peak structure. Such results are usually explained by a dynamical bifurcation\cite{PhysRevB.81.144518,PhysRevB.76.014524}. Further analysis of this phenomenon is beyond the scope of the present work, but it is reassuring that our simple model also gives rise to this complex behaviour.

\section{Driving with Constant Bias Current}
\label{sec:const_bc}
We now maintain the bias current at a constant, $I=0.96$, close to the switching value, and we apply a microwave field to the junction. As we have already seen, the tunnelling rate is increased if the resonance condition is met. In Fig. \ref{fig:switch_frequency} we observe what happens at different frequencies. First, the switching probability follows the field strength, but as we approach resonance, a peak emerges, and when the frequency is increased even further the peak is reduced. In the color plot we see a hyperbolic behaviour of the signal following each oscillation cycle, and for increasing frequency they, naturally, move closer together and increase in strength and as a consequence the later ones disappear. After the resonance we see that the signal slowly smears out.

A switching rate that follows the field strength in real time allows implementation of a field amplitude detector capable of resolving single oscillations of the field. In Fig. \ref{fig:switch_amplitude} we increase the field strength while maintaining a very low frequency and see that when the field is strong enough single positive oscillations of the field amplitude are clearly detected. In the color plot we also clearly see the growth and reduction of equally spaced areas of probability. 

We intend to study the performance and sensitivity of this detection mode of the CBJJ further with particular emphasis on the prospects for quantum information processing in circuit quantum electrodynamics.

\section{Conclusion and Outlook }
\label{sec:conclusion}

We have introduced a novel, effective method using time dependent imaginary potentials to describe the switching behaviour of a current biased Josephson junction. The method is readily implemented with standard wave packet solvers and with realistic parameters, it  reproduces a wide range of results that have previously been investigated experimentally and theoretically by other techniques \cite{PhysRevB.81.144518, PhysRevB.77.104531, PhysRevLett.90.037003, PhysRevB.68.060504, PhysRevB.81.144518, PhysRevB.76.014524,Davis06freq,PhysRevLett.81.2679}. We emphasize that our treatment builds on an ansatz for a time dependent imaginary potential (TDIP), and a number of possibilities may be explored for quantitative improvement of the potential chosen.

An advantage of the TDIP method is its ability to deal with explicitly time dependent driving fields. With the explicit treatment of the wave function of the phase variable conditioned upon the tunneling dynamics, we may readily extend the studies towards other systems where non-trivial tunneling dynamics in time-dependent potentials is present. We may also expect that the methods developed here can be used to model the dynamics of coupled CBJJs and of correlated and entangled CBJJ dynamics with the quantized radiation fields. Potentially our approach may thus form the basis for a novel quantum theory of measurement, where the noise and back action is adapted to the special field amplitude sensitivity of the device.

\section{Acknowledgement}
The authors acknowledge support from the EU 7th Framework Programme collaborative project iQIT.

\bibliography{bt}

\end{document}